\documentclass[a4paper,11pt]{article}
\usepackage{jheppub} % for details on the use of the package, please
\usepackage[normalem]{ulem} %for \sout comment

\def\be{\begin{eqnarray}}
\def\ee{\end{eqnarray}}

\newcommand{\dyc}[1]{\textcolor{black}{ #1}}

\usepackage{comment}

\def\be{\begin{eqnarray}}
\def\ee{\end{eqnarray}}

\title{Effective Theory Approach for Axion Wormholes}

\preprint{CERN-TH-2023-184}

\author[a,b]{Dhong Yeon Cheong,}
\author[a, e, 1]{Seong Chan Park,\note{Corresponding author.}}
\author[c,d,e,1]{and Chang Sub Shin }

\affiliation[a]{Department of Physics and IPAP, Yonsei University, \\ Seoul 03722, Republic of Korea}
\affiliation[b]{Theoretical Physics Department, CERN, \\ 1211 Geneva 23, Switzerland}
\affiliation[c]{Department of Physics and Institute of Quantum Systems (IQS),
Chungnam National University, \\ Daejeon 34134, Republic of Korea}
\affiliation[d]{Center for Theoretical Physics of the Universe, Institute for Basic Science (IBS), \\ Daejeon, 34126, Korea}
\affiliation[e]{Korea Institute for Advanced Study, \\Seoul 02455, South Korea}

\emailAdd{dhongyeon@yonsei.ac.kr}
\emailAdd{sc.park@yonsei.ac.kr}
\emailAdd{csshin@cnu.ac.kr}

\abstract{
We employ the effective field theory approach to analyze the characteristics of Euclidean wormholes within axion theories. 
Using this approach, we obtain non-perturbative instantons in various complex scalar models with and without a non-minimal coupling to gravity, as well as models featuring the $R^2$ term for a range of coupling values.
This yields a series of analytical expressions for the axion wormhole action, shedding light on the model parameters and field dependencies of contributions in both the ultraviolet and infrared domains. Consequently, model-dependent local operators that disrupt axion shift symmetries are generated at lower energy levels. This, in turn, provides crucial insights into the gravitational influences on the axion quality problem.
}

\begin{document} 

\maketitle
\flushbottom

\section{Introduction}
\label{sec:intro}

The axion, a pseudo-Nambu-Goldstone Boson of a global $U(1)$ symmetry, provides a novel solution to the strong CP problem~\cite{Peccei:1977hh,Peccei:1977ur, Wilczek:1977pj, Weinberg:1977ma}, and is associated with various phenomenological puzzles, %including 
such as the dark matter problem ~\cite{Preskill:1982cy, Abbott:1982af, Dine:1982ah}, dark energy problem or both (see \cite{Kim:2008hd, Marsh:2015xka, DiLuzio:2020wdo, Choi:2020rgn,Ferreira:2020fam} for recent reviews). 
However, gravity may jeopardize these solutions, as no global symmetry is allowed in a gravitating system~\cite{Hawking:1987mz, Giddings:1988cx, Banks:2010zn, Witten:2017hdv, Harlow:2018jwu}.

The challenge of preserving the shift symmetry of the axion, the so-called Peccei-Quinn (PQ) symmetry $U(1)_\text{PQ}$, against numerous potential sources of global symmetry-breaking has been coined as the axion quality problem~\cite{Dine:1986bg,Kamionkowski:1992mf, Barr:1992qq, Holman:1992us, Ghigna:1992iv}.
Many attempts have been made to resolve this issue~\cite{Witten:1984dg,Randall:1992ut,Cheng:2001ys,Izawa:2002qk,Choi:2003wr, Lillard:2018fdt, Ardu:2020qmo, Alvey:2020nyh, Hamaguchi:2021mmt, Dvali:2022fdv, Cheong:2022ikv, Bonnefoy:2022vop, Burgess:2023ifd, Choi:2023gin}, but subtleties still exist as quantum gravitational contributions are largely unknown and non-perturbative in their natures.  

One particularly important direction to calculate a non-perturbative contribution is to estimate the effect of the Euclidean axion wormhole, the stationary point solution for the Euclidean action with gravity whose throat is connected to an asymptotically flat region where the global PQ charge flows~\cite{Lee:1988ge, Giddings:1989bq, Abbott:1989jw, Coleman:1989zu, Kallosh:1995hi, Hebecker:2016dsw,  Alonso:2017avz, Hertog:2018kbz, Hebecker:2018ofv, Loges:2022nuw, Andriolo:2022rxc, Loges:2023ypl,Jonas:2023ipa}.
If the wormhole action  $S_{wh}$ is large enough, a semi-classical approximation can be made without needing a complete UV theory of gravity.  As the gravitational impact on IR physics is suppressed by an exponential factor $e^{-S_{wh}}$,  to a good approximation, $S_{wh}\gtrsim 200$ guarantees the axion quality~\cite{Kallosh:1995hi}. An early attempt by Giddings and Strominger~\cite{Giddings:1989bq} suggested $S_{wh}\sim M_P/f_a$, thus, $f_a \lesssim M_P/200$ seemed to define the preferred parameter region for the axion quality, however, it turned out that the dynamics of the accompanying radial scalar partner of the axion completely modifies the conclusion, as $S_{wh}\sim \log M_P/f_a$~\cite{Lee:1988ge,Abbott:1989jw, Coleman:1989zu,  Kallosh:1995hi}.

Recently, it is noticed that a large non-minimal coupling of the complex scalar to the Ricci scalar ($\xi \Phi^* \Phi R$) can make the wormhole action large enough in both metric and Palatini formalisms of gravity~\cite{Hamaguchi:2021mmt,Cheong:2022ikv}.
Here, the complex scalar $\Phi=\frac{1}{\sqrt{2}}\phi e^{i\theta}$ has an axion as its 
complex phase field and its radial partner.
The non-minimal coupling term naturally arises in an effective theory or is radiatively generated ~\cite{Ford:1981xj}, and has been actively investigated for inflationary cosmology in metric, Palatini or both formulations~\cite{Futamase:1987ua, Bezrukov:2007ep, Park:2008hz, 
Bezrukov:2010jz, Hamada:2014wna, Hamada:2014iga, Jinno:2019und}. 

On the other hand, both studies in Ref.~\cite{Hamaguchi:2021mmt,Cheong:2022ikv}  heavily rely on numerical calculations
and still need an analytic understanding of the full parameter space. 
Therefore, in this paper, we enlarge our scopes and attempt to study the analytical structure of these Euclidean wormholes. 
In particular, we take an effective field theory approach by imposing boundary conditions of wormholes with {\bf IR} field degrees of freedom,  along with identifying the generation of associated PQ breaking operators at low energies.  
This approach dramatically simplifies our analysis and allows us to obtain an analytic expression of the wormhole action that is solely determined by the {\bf IR} fields. The solution provides a transparent understanding of the parameter dependence of the action and perspectives on generalizations with possible limitations.

This paper is organized as follows. We first present the generic Lagrangian structure associated with axion theories, and take the effective theory approach to obtain a generic formalism representing the Euclidean wormhole solutions. We demonstrate that the wormhole action can be decomposed to its UV and IR components, with the IR component containing the UV cutoff of the theory. We then demonstrate this formalism in several different examples, each representing a different class of models. We conclude with the implications of this formalism.

\section{Eulcidean Wormhole for General Axion Models}

\subsection{General Lagrangian}

We introduce a set of axion fields $\{\theta\} =\{\theta^I(x)\}_{I=1,2,\cdots,n_a}$, characterized as `angular fields' due to their periodic and compact nature, adhering to the discrete gauge symmetries $\theta^I(x)\equiv \theta^I(x)+2\pi$. Alongside these symmetries, the axions exhibit continuous global symmetries, known as the PQ symmetry, described by the transformation
\begin{align}
	U(1)_{\rm PQ}^I :  \theta^{I} \rightarrow \theta^{I} + c^{I}\,,~~c^{I} \in \mathbb{R} .
\end{align}
In conjunction with the axions, we introduce `radial fields', $\{\phi\}=\{\phi^A(x)\}_{A=1,2,\cdots, n_s}$, whose vacuum expectation values (VEVs) determine the axion decay constants. These dynamic scalar fields are commonly observed in various axion scenarios, including UV completions of the PQ symmetry~\cite{Abbott:1989jw, Kallosh:1995hi, Alonso:2017avz, Alvey:2020nyh}.
Notably, the number of radial scalars can differ from the count of the axions, i.e., $n_s \neq n_a$.
The general action for the axions and their radial fields in the Einstein frame is  given by 
\begin{align} \label{eq:general action0}
\mathcal{S}= \int d^4 x \sqrt{-g} \left(-\frac{M_P^2}{2} R + \frac{1}{2} 
\sum_{A,B}^{n_s} G_{AB}(\phi)\partial_\mu \phi^A \partial^\mu \phi^B + \frac{1}{2}  \sum_{I,J}^{n_a}( f^2_a(\phi))_{IJ} \partial_\mu\theta^I \partial^\mu\theta^J\right), 
\end{align}
where $M_P=1/\sqrt{8\pi G} \approx 2.4\times 10^{18} ~{\rm GeV}$ denotes the reduced Planck mass,  $G_{AB}(\phi)$ and $(f^2_a(\phi))_{IJ}$ are field dependent metrics in the field spaces of the radial scalar fields and the axion fields, respectively. 
Implicitly, we assumed that a potential for the radial fields provides the vacuum expectation values of $\phi^A$ as $\langle\phi^A\rangle$ 
so that $(f_a^2(\langle \phi\rangle))_{IJ}= (f_a^2)_{IJ}$; however, its explicit form is not relevant for our discussions in the IR region~\cite{Abbott:1989jw, Coleman:1989zu}. 
\dyc{We discuss the effect of potential terms more explicitly in Appendix~\ref{App:potential}.}
The CP properties of the axions and the radial fields are given as
\begin{align*}
CP: (\theta^I, \phi^A)\to (-\theta^I, \phi^A), ~~{}^\forall I, A,
\end{align*}
thus no kinetic mixing is allowed between any axion and raidal field.

The discrete gauge symmetry of the axions becomes more transparent by dualizing the axion field to a three form field strength $H_{I\mu\nu\rho}$: 
\begin{align}
\mathcal{S} &= \int d^4 x \sqrt{-g} \left(-\frac{M_P^2}{2} R+ \frac{1}{2} G_{AB} (\phi) \partial_\mu\phi^A \partial^\mu \phi^B \right. \nonumber\\
& \hskip 2.5cm \left.  +\, \frac{1}{12} (f^{-2}_a(\phi))^{ IJ} H_{I\mu\nu\rho} H^{\mu\nu\rho}_J  + \frac{1}{6} \theta^I  \epsilon^{\mu\nu\rho\sigma} \partial_\mu H_{I \nu\rho\sigma} \right), 
\end{align}
where the covariant Levi-Civita tensor is given by $\epsilon_{\mu\nu\rho\sigma} =  (-1)^{\pi(\mu\nu\rho\sigma)}\sqrt{-g}=-(\epsilon^{\mu\nu\rho\sigma})^{-1}$ for the signature of the permutation $\pi(\mu\nu\rho\sigma)$. 
Using the differential form notation, the equations of motion for $\theta^I$ give $H_I =(1/6)H_{I\mu\nu\rho}dx^\mu\wedge dx^\nu\wedge dx^\rho = dB_I$, where $B_I=(1/2) B_{I\mu\nu} dx^\mu\wedge dx^\nu$ is the 2-form gauge field. The equations of motion for $H_I$ give $H_I = (f^2_a(\phi))_{IJ} \star d\theta^J$.  Inserting these solutions of $H_I$ to the action, we recover the action for the axions in Eq.~(\ref{eq:general action0}).

 For the wormhole solution, we derive the Euclidean action after Wick rotation 
\begin{align}\label{eq:general action}
\mathcal{S}_E =& \int d^4 x \sqrt{g^E} \left(-\frac{M_P^2}{2} R_E+ \frac{1}{2} G_{AB} (\phi) \partial_\mu\phi^A \partial^\mu \phi^B + \frac{1}{12} (f^{-2}_a(\phi))^{ IJ} H_{I\mu\nu\rho} H^{\mu\nu\rho}_J   \right) \nonumber\\ 
&  - i  \int d^4 x   \sqrt{g^E}  \left(\frac{1}{6} \theta^I \epsilon_E^{\mu\nu\rho\sigma} \partial_\mu H_{I \nu\rho\sigma} \right) ,
\end{align}
where the Euclidean metric is $g^E_{\mu\nu} = \delta_{\mu\nu}$ in the limit of flat space, the Euclidean curvature scalar is $R_E$, and the Euclidean Levi-Civita tensor is given by $\epsilon^E_{\mu\nu\rho\sigma} = (-1)^{\pi (\mu\nu\rho\sigma)}\sqrt{g^E}=(\epsilon_E^{\mu\nu\rho\sigma})^{-1}$.  
\begin{figure}[t]
	\centering
	\includegraphics[width=.5\textwidth]{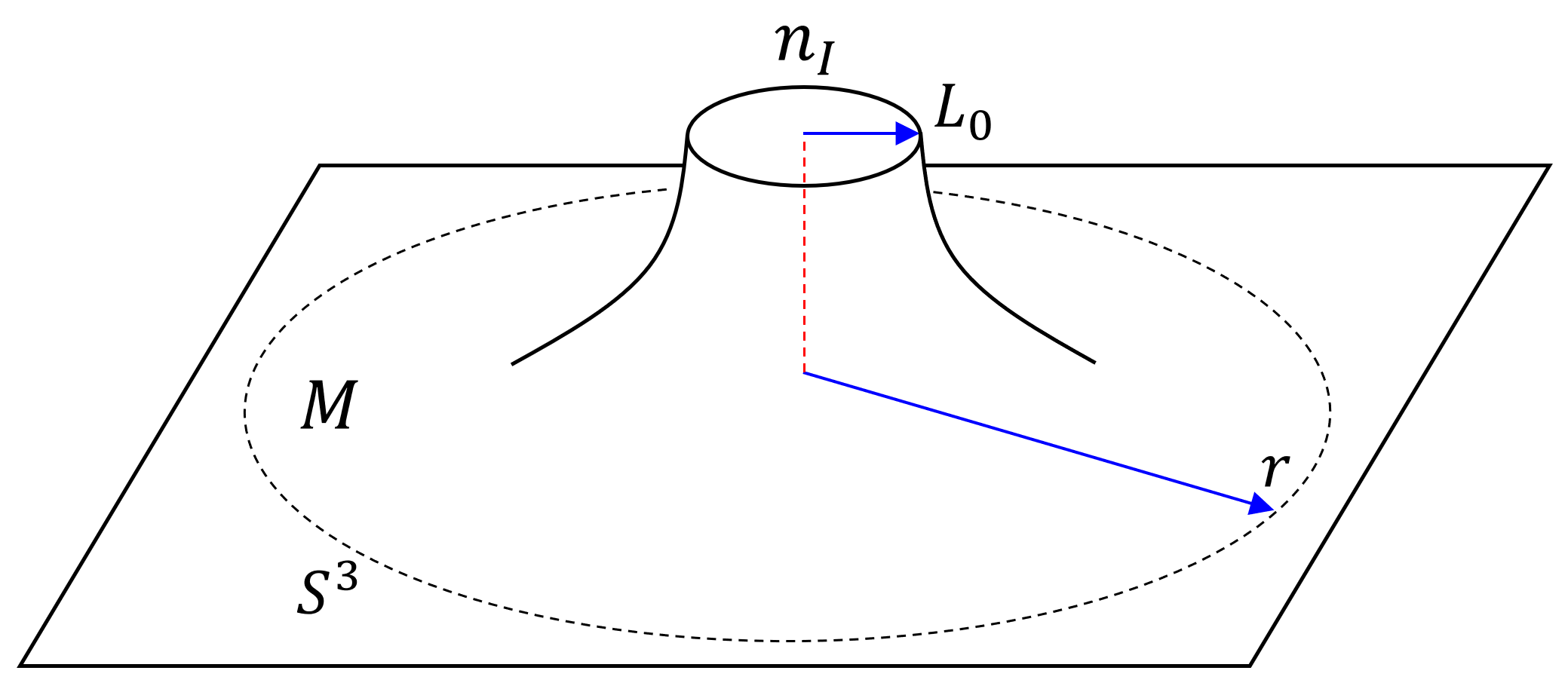} % requires the graphicx package
	\caption{A schematic diagram of the Euclidean wormhole geometry}
	\label{fig:wormhole_geometry}
\end{figure}
The geometry of the Euclidean wormhole  is solved by varying $\delta S_E$: 
\begin{align} \label{eq:wormhole_geometry}
	ds^2_{wh} = g_{rr}^E dr^2 + r^2d\Omega_3^2= \frac{dr^2}{\left(1 - L_0^4/r^4 \right)} + r^2 d\Omega_{3}^2,
\end{align}
where $L_{0}$ denotes the size of the wormhole throat radius. 
The schematic shape of the Euclidean wormhole is depicted in Fig.~\ref{fig:wormhole_geometry}.
It is intriguing to notice that the Ricci scalar in the geometry Eq.~(\ref{eq:wormhole_geometry}) is solely determined by $L_0$ and the distance from the center of the wormhole
\begin{align} \label{eq:Ricci_scalar}
	R_E = - \frac{6L_{0}^4}{r^6}, ~~~~~L_0 < r < \infty.
\end{align}
Effectively, the `interior' of the wormhole is in $r<L_0$ and provides a source of nonzero, integer Peccei-Quinn charge:
\begin{align} \label{eq:PQcharge}
	\int_{\partial M} H_I = n_I \in \mathbb{Z},
\end{align}
where $\partial M\sim S^3$ is the boundary of the wormhole. 

Finally, the Euclidean action Eq.~(\ref{eq:general action}) with the wormhole geometry is given as 
\begin{align}
	S_E[\textrm{wormhole}] = S_{wh}[n, \phi] - i n_I \theta^I
\end{align} for the background axion fields $\theta^I$. Here, the first term $S_{wh}[n, \phi]$ is from the RHS in the first line of Eq.~(\ref{eq:general action}), and the second term $-i n_I \theta^I$ is from the RHS in the second line.\footnote{Actually, $S_{wh}$ here is the half-wormhole action (or instanton action) without the Gibbons-Hawking-York (GHY) boundary term~\cite{York:1971hw,York:1972sj,Gibbons:1976ue}. 
The GHY surface term is less important in our discussion, so we will not discuss its contribution explicitly. To make the discussion more concise, we also use the term ``wormhole action" to represent $S_{wh}[n,\theta]$ in the whole discussion.}
The saddle point approximation in the path integral 
yields PQ breaking local operators 
that are suppressed by non-perturbative instanton factor 
\begin{align} 
	e^{-S_E[\textrm{wormhole}]} = & e^{-S_{wh}[n, \phi] + i n_I \theta^I}.
\end{align}  
One can notice that the $2\pi$-periodicity of the axions provides the Dirac quantization condition for the PQ charge.

\subsection{Field dependent axion wormhole solutions}
For the metric ansatz of the Euclidean wormhole Eq.~(\ref{eq:wormhole_geometry}),
the field profiles are also evaluated as the $O(4)$ symmetric solutions: $\phi^A(r)$ and $H_{I\mu\nu\rho}(r)$. First of all,  with the quantization condition Eq.~(\ref{eq:PQcharge}), the equation of motion for $H_{I\mu\nu\rho}$ gives the solution
\begin{align} \label{eq:PQ-charge_sol}
 H_I =  n_I  \star  \frac{dr}{2\pi^2}.
\end{align}
Next, one can obtain the equations of motion for metric and scalar fields with respect to the distance $r\in [L_0, \infty)$ after inserting the solution of $H_I$. 
Instead of using $r$, it is more convenient to take a new variable $\tau$ defined as \cite{Arkani-Hamed:2007cpn} \begin{align}
	\tau(r)=\frac{1}{4\pi^2L_0^2} \arctan\left(\sqrt{r^4/L_0^4-1}\right). 
\end{align}
The range of $\tau$ is from $\tau(L_0)=0$ to $\tau(\infty)= 1/8\pi L_0^2=\tau_\infty$. 
The wormhole action $S_{wh}$ includes the Einstein-Hilbert action, the kinetic term of the radial scalars, and 
the kinetic term for the Euclidean 2-from gauge fields $B_{I\mu\nu}$ with the quantization condition, Eq.~(\ref{eq:PQ-charge_sol}). With the  integral variable $\tau$, $S_{wh}$ can be written as 
\begin{align}
S_{wh}=	\int_0^{\tau_\infty} d\tau \left( 12\pi^4  M_P^2 L_0^4 
	+ \frac{1}{2} G_{AB}(\phi)\frac{d\phi^A}{d\tau}\frac{d\phi^B}{d\tau} 
	+ \frac{1}{2}(f^{-2}_a(\phi))^{IJ} n_I n_J \right).
\end{align}
The equations of motion for $\phi^A$ give
\begin{align} \label{eq:eom_for_phi1}
	G_{AB}(\phi) \frac{d^2\phi^B}{d\tau^2} +  \Gamma_{ABC}(\phi) \frac{d\phi^B}{d\tau} \frac{d\phi^C}{d\tau} = \frac{\partial}{\partial\phi^A}\left(\frac{1}{2} (f^{-2}_a(\phi))^{IJ}n_I n_J\right),
\end{align}
where the `connection' in the field space is 
\begin{align}
	\Gamma_{ABC}(\phi) = \frac{1}{2} \left(\frac{\partial G_{AB}(\phi)}{\partial\phi^C} +\frac{\partial G_{AC}(\phi)}{\partial \phi^B} - \frac{\partial G_{BC}(\phi)}{\partial \phi^A} \right).
\end{align}
The trace for Einstein's equation gives 
\begin{align}\label{eq:eom_for_phi2}
\dyc{ 12\pi^4 M_P^2 L_0^4 =  \frac{1}{2} (f^{-2}_a(\phi))^{IJ} n_I n_J -\frac{1}{2} G_{AB}(\phi)\frac{d\phi^A}{d\tau}\frac{d\phi^B}{d\tau}  .}
\end{align}
For given {\bf IR} degrees of freedom $\phi^A$, 
the boundary conditions are given by  
\begin{align} 
	\left.\frac{d\phi^A(\tau)}{d\tau}\right|_{\tau=0}=0,\quad 
	\phi^A(\tau_\infty) =\phi^A.
\end{align} 
In general, the throat field values $\phi_0=\phi(\tau=0)$ are functions of {\bf IR} fields, with its sensitivity being quite model dependent.
The throat radius is directly given by the axion decay constant at the throat for given wormhole charges $n_I$,
\begin{align} \label{eq:throat_radius}
	12\pi^4 M_P^2L_0^4  = \frac{1}{2} (f^{-2}_a(\phi_0))^{IJ}n_In_J.
\end{align} 
One can easily notice that Eq.~(\ref{eq:eom_for_phi2}) with Eq.~(\ref{eq:throat_radius})
 resemble equations in classical mechanics describing the motion of point particles. Using the canonical momenta $p_A$ and the potential $V$ of $\phi$ given by
\begin{align}
p_A = G_{AB}(\phi)\frac{d\phi^B}{d\tau}, \quad  V = \frac{1}{2} (f^{-2}_a(\phi_0))^{IJ}n_I n_J -  \frac{1}{2} (f^{-2}_a(\phi))^{IJ}n_I n_J,
\end{align}
 Eq.~(\ref{eq:eom_for_phi1}) can be written as 
\begin{align} \label{eq:eom_for_scalars}
     \frac{dp_A}{d\tau} + \frac{1}{2}\left(\partial_A G^{BC}\right) p_B p_C   + \partial_A V =0 ,\quad
     \frac{1}{2} G^{AB} p_A   p_B  +V = 0,
\end{align}
where $G^{AB}G_{BC}={\delta^A}_C$. The potential $V(\phi)$ and the metric of the kinetic terms $ G_{AB}(\phi)$ do not explicitely depend on $\tau$, so if the solution exists, $p_A$s are given by the functions of $\phi^A$s, i.e.  $p_A(\phi)$.
For given {\bf IR} field values $\phi$, the throat values $\phi_0$ are determined by the equations,
\begin{align} \label{eq:boundary_conditions}
\sqrt{(f^{-2}_a(\phi_0))^{IJ}n_I n_J} \int_{\phi_0^A}^{\phi^A}     \frac{d\varphi^A}{M_P} \frac{1}{G^{AB}(\varphi)p_B(\varphi)} = 
\frac{\pi\sqrt{6}}{4}   \quad (\textrm{for each $A$}),
\end{align}
Utilizing the equations of motion,
the wormhole action can be represented by several ways.  
\begin{align}\label{eq:wormhole_action}
	S_{wh}[n, \phi] &=  \int_0^{\tau_\infty} d\tau  (f^{-2}_a(\varphi))^{IJ}n_In_J =  3\pi^3  M_P^2 L_0^2  + \int_{\phi_0}^\phi  d\varphi^A\, p_A(\varphi) \nonumber\\
	 &=  \frac{\pi \sqrt{6}}{4} M_P  \sqrt{(f^{-2}_a(\phi_0))^{IJ}n_I n_J } 
+ \int_{\phi_0}^\phi d\varphi^A\, p_A(\varphi).
\end{align}
The last expression of Eq.~(\ref{eq:wormhole_action}) shows how $S_{wh}$ consists of the sum of the UV (near the throat) and IR (the field dependent) contributions. 
The first term of the RHS corresponds to the form of the Giddings-Strominger wormhole~\cite{Giddings:1989bq},
 and the second term represents the effect of the radial field's profile in the wormhole geometry. 
 It contributes to the action positively and determines how 
 the wormhole action depends on the {\bf IR} field degrees of freedom. 
   
 { Before moving on to the next section, we outline our assumptions for evaluating the wormhole action. First, we take the Planck scale as the cut-off scale of the 4-dimensional effective field theory. The actual cut-off scale $\Lambda$ of the effective theory may be much lower than the Planck scale. If the throat radius $L_0$ in our evaluation is less than $1/\Lambda$, the effects induced by operators suppressed by the cut-off scale should not be ignored, which could invalidate the existence of wormhole solutions. Therefore for a valid wormhole solution to exist, dimensional analysis requires $L_0 \gtrsim 1/\Lambda$, leading to a wormhole action of order $M_P^2/\Lambda^2$, which can be easily ${\cal O}(100)$ for $\Lambda\lesssim {\cal O}(0.1)\,M_P$.  In this sense, the axion quality problem may imply a low cutoff scale for the four-dimensional effective theory.

 However, for the contribution of ${\cal O}(M_P^2/\Lambda^2)$ to the action $S_{wh}$, the prefactor  heavily relies on UV physics. Therefore, as a conservative approach, we explicitly evaluate the axion wormhole solution for various models with minimal assumption about new physics at the UV and see the validity of our approach. For instance, if the value of the wormhole action is not sufficiently larger than ${\cal O}(1)$, we conclude that the axion quality issue is highly dependent on the UV physics such as the cut-off scale, higher dimensional operators, multiple wormhole contributions, etc. On the other hand, if the wormhole action is much larger than ${\cal O}(1)$, we consider that the solution is less sensitive to UV physics, providing a feasible solution to the axion quality problem. In this study, we will take a more conservative view of the cut-off scale and the higher dimensional operators and focus on clarifying the analytic structure of the wormhole action. Based on the solutions, we investigate the conditions under which the wormhole action can be sufficiently large.}

\section{Case Studies: Single Axion Models}
\subsection{Complex scalar model}
The structure of the wormhole action becomes more transparent in single axion models. 
We start with the minimal field content, a single axion plus a radial mode associated with a single $U(1)_\text{PQ}$.  In this case, we take $\phi^1=\phi$, $\theta^1=\theta$, and  $G_{11}=G(\phi)$, $(f^2_a(\phi))_{11} = f^2_a(\phi)$.  The action takes the form  
\begin{align} \label{eq:single_axion_model}
	\mathcal{S} = \int d^4 x \sqrt{-g} \left( - \frac{M_P^2}{2} R + \frac{1}{2}    G(\phi)\partial_{\mu} \phi \partial^{\mu} \phi  + \frac{1}{2}  f^2_a(\phi)  \partial_{\mu} \theta  \partial^{\mu} \theta  \right) .
\end{align} 
Using the master formulas Eqs.~(\ref{eq:throat_radius}) and (\ref{eq:eom_for_scalars}), we obtain  the throat radius $L_0$ and the scalar field equation of motion as 
\begin{align} \label{eq:eom for phi_single}
	L_0^2  =\frac{1}{2\pi^2\sqrt{6}}\left(\frac{n}{M_P f_a(\phi_0) }\right), \quad
	 p^2(\phi) =  G(\phi) \left(\frac{n^2}{f^2_a(\phi)} -\frac{n^2}{f^2_a(\phi_0)}\right).
\end{align}
Using Eq.~(\ref{eq:boundary_conditions}) we obtain the condition for the throat value of $\phi_0$ as the function of  \textbf{IR} field $\phi$.
\begin{align} \label{eq:boundary condition_single}
\int_\phi^{\phi_0} \frac{d\varphi}{M_P}\,  \frac{\sqrt{G(\varphi)}}{\sqrt{f^2_a(\phi_0)/f^2_a(\varphi) -1}} =  
\frac{\pi \sqrt{6}}{4}.
\end{align}
Note that the solution $\phi_0$  is independent of $n$, while the radial profile of $\phi(\tau)$ depends on $n$ as Eq.~(\ref{eq:eom for phi_single}).
The corresponding wormhole action Eq.~(\ref{eq:wormhole_action}) becomes 
\begin{align}  \label{eq:wormhole_action_single_axion}
	S_{wh}[n,\phi] & =\int_0^{\tau_\infty} d\tau \frac{n^2}{f^2_a(\varphi)}=   
	\int_{\phi_0}^{\phi} \frac{d\varphi}{p(\varphi)} \frac{G(\varphi)n^2}{ f^2_a(\varphi)}
 \nonumber\\
	 & =  n \int_\phi^{\phi_0} \frac{d\varphi}{f_a(\varphi)} \frac{\sqrt{G(\varphi)}}{\sqrt{1- f^2_a(\varphi)/f^2_a(\phi_0)}}.
\end{align}
In the second line of the equation, we assume $\phi_0> \phi$. 
Because $\phi_0$ is independent of $n$, the wormhole action is linearly dependent on the wormhole charge $n$.   

In the following sections, we derive several analytic results for the wormhole action in various models of single axion cases.

\subsubsection{Model with a non-minimal coupling to gravity} 

We consider an example of a complex scalar field. Here we allow a non-minimal coupling to gravity~\cite{Coule:1989xu, Coule:1992pz}. 
The action is 
\begin{align}\label{eq:complexscalar_model}
	\mathcal{S} = \int d^4 x \sqrt{-g} \left[ - \left(\frac{M_P^2}{2} + \xi |\Phi|^2\right) R +     \partial_{\mu} \Phi \partial^{\mu} \Phi^*  \right] .
\end{align}
Decomposing the complex scalar into the radial mode $\phi$, and the angular mode (axion) $\theta$,
\begin{align}
	\Phi(x)=\frac{1}{\sqrt{2}}	\phi(x) e^{i\theta(x)} 
\end{align}
the action in the Einstein frame becomes 
\begin{align}  \label{eq:complexscalar_Einsteinframe}
\mathcal{S} = \int d^4 x \sqrt{-g} \left( -\frac{M_P^2}{2} R 
+ \frac{1}{2}   G(\phi) \partial_\mu\phi\partial^\mu\phi +\frac{1}{2}   f^2_a(\phi) \partial_\mu\theta \partial^\mu \theta\right),
\end{align} 
where 
\begin{align} \label{eq:kinetic_metric}
	G(\phi)  = \frac{ M_P^2(M_P^2 + \xi \phi^2(1+ \alpha \xi))}{( M_P^2 +\xi \phi^2)^2},\quad
	f_a(\phi)  = \frac{M_P \phi}{\sqrt{M_P^2 +\xi\phi^2}}. 
\end{align}
We introduce a constant $\alpha$ that represent different gravity formalisms, 
\begin{align}
	\alpha  &= 6\quad \textrm{(Metric formalism)}  \nonumber \\
	        &= 0 \quad \textrm{(Palatini formalism)}   
\end{align}
Following the equations of motion for Euclidean action, 
we get two master formulae for the boundary conditions and the wormhole action.
For boundary conditions, Eq.~(\ref{eq:boundary condition_single}) gives 
\dyc{\begin{align}
&\sqrt{\frac{(1+\alpha\xi)(1+\xi \phi_0^2/M_P^2)}{\xi}} \arccos\sqrt{\frac{1+\xi \phi^2(1+\alpha\xi)/M_P^2}{1+ \xi\phi_0^2(1+\alpha\xi)/M_P^2}} \nonumber\\
&-\sqrt{\alpha} \arctan \sqrt{\frac{(\phi_0^2-\phi^2)\alpha\xi^{2}/M_P^2}{ (1+\xi\phi^2(1+\alpha\xi)/M_P^2)(1+\xi\phi_0^2/M_P^2)}} = \frac{\pi\sqrt{6}}{4}.
\end{align}}
We expect that $\phi_0$ is proportional to the Planck mass $M_P$. Therefore, one can safely take an approximation of $\phi\ll \phi_0$, where $\phi$ is the {\bf IR} field whose actual value is hierarchically smaller than the cut-off scale of the model. Ignoring the ${\cal O}(\phi/\phi_0)$ contributions, $\phi_0$ is provided by
\begin{align} \label{eq:phi0}
		\frac{\phi_0}{M_P} &= \frac{\pi\sqrt{6}}{4}  \left(1 + {\cal O}(\xi) \right) \hskip 2cm (\xi\ll 1) \nonumber\\
	  &= \sqrt{\frac{5}{4\hskip 0.04cm\xi}}   \left(1  + {\cal O}(1/\xi )\right)  \hskip 1.6cm  (\xi \gg 1,\ \alpha=6) \nonumber\\
	  &= \sqrt{\frac{3}{2}} \left(1  + {\cal O}(1/\sqrt{\xi})\right)  \hskip 1.25cm  (\xi\gg1,\ \alpha=0) .
\end{align}
We note that the throat field value $\phi_0$ in the Palatini formalism is different from that in the metric formalism.  However, the wormhole throat size is insensitive to $\phi_0$ because the axion decay constant $f_a(\phi)$ becomes independent of $\phi$ as $\xi\phi^2\gtrsim M_P^2$, and converges to the form  $M_P/\sqrt{\xi}$. The throat radius is given by 
\begin{align} \label{eq:L0sol}
	L_0^2 = \frac{1}{2\pi^2\sqrt{6}}\left(\frac{n}{ M_P f_a(\phi_0) } \right) &= \frac{n}{3\pi^3}\left(\frac{1}{M_P^2} \right) \hskip 1.7cm (\xi\ll 1)   \nonumber\\
	&=\frac{n\sqrt{3}}{2\pi^2\sqrt{10}} \left(\frac{\sqrt{\xi}}{M_P^2}\right) \hskip 1cm (\xi\gg 1,\ \alpha=6)\nonumber\\ 
	&= \frac{n}{2\pi^2\sqrt{6}} \left(\frac{\sqrt{\xi}}{M_P^2}\right) \hskip 1.2cm (\xi\gg 1, \ \alpha=0)  .
\end{align}
Finally, the corresponding wormhole action becomes
\begin{align}
	S_{wh}[n,\phi] &=n \left[
	 \sqrt{\Delta_{\xi,\alpha}} \arccos\Big(\frac{1}{\sqrt{1+\Delta_{\xi,\alpha}}}\Big) + \ln\frac{2\phi_0}{\phi\sqrt{1+ \Delta_{\xi,\alpha}}} \right],
\end{align} where 
$\Delta_{\xi, \alpha} = \xi\phi_0^2(1+\alpha\xi)/M_P^2$. 
\begin{figure}[t]
	\centering
	\includegraphics[width=.48\textwidth]{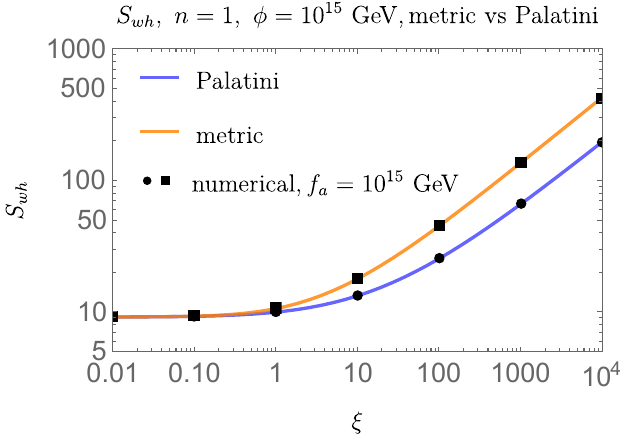} \includegraphics[width=.48\textwidth]{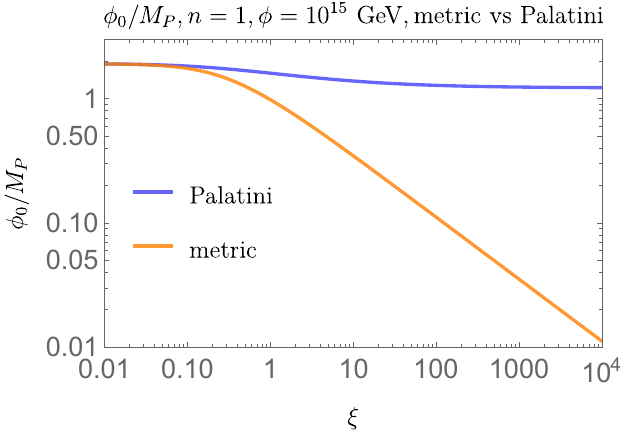} % requires the graphicx package
	\caption{(Left) The full wormhole action $S_{wh}$ for a complex scalar with a non-minimal coupling for both formalisms. The solid lines correspond to the full analytic computations with $\phi=10^{15}\,{\rm GeV}$, and the numerically obtained values correspond to the markers with $f_a=10^{15}\,{\rm GeV}$ as in Ref.~\cite{Cheong:2022ikv}. It clearly shows the behavior of the action $S_{wh} \sim \sqrt{\xi}$ for $\xi\gg 1$. (Right) The \textbf{UV} field value $\phi_0$ for both metric and Palatini scenarios. The values correspond to the minimal coupled case in the $\xi\ll 1$ regime, whereas in the $\xi \gg 1 $ regime the behaviors deviate, with the Palatini case saturating and the metric case decreasing as $1/\sqrt{\xi} $  }
	\label{fig:nonminimalaction_compare}
\end{figure}

\dyc{
Fig.~\ref{fig:nonminimalaction_compare} compares the action results from our effective theory approach with $\phi=10^{15}\,{\rm GeV}$ with the actions calculated numerically using a shooting method implemented in~\cite{Cheong:2022ikv}. We use a potential 
\begin{align}
    V(\phi) = \frac{\lambda }{4} \left(\phi^2 - f_{a}^2  \right)^2
\end{align}
that gives the vacuum value of the axion decay constant $f_a=10^{15}\, {\rm GeV}$. } It clearly shows the validity of our approximation in which the parameter dependence becomes transparent. 
Depending on the values of $\xi$ and $\alpha$, the form of the wormhole action can be decomposed into the UV part mostly given by the throat regime, and the IR part logarithmically dependent on the radial field $\phi$:
\begin{align}
	S_{wh}[n,\phi] & \simeq   n \left(S_{wh}^{\rm UV}[\xi] +  \ln\frac{\Lambda_{\rm eff}[\xi]}{\phi}\right).
  \label{eq:actionnonminimal}
\end{align}  
The UV part has the form $S_{wh}^{\rm UV}[\xi] = c_0 + c_1 \sqrt{\xi}$:
\begin{align}
 S_{wh}^{\rm UV}[\xi] &=  \ln \frac{\pi\sqrt{6} }{2}\hskip 3.cm  {\rm for}\  \xi\ll 1
	\nonumber\\
	& = \ln\frac{\sqrt{6}}{3} +   \frac{\pi \sqrt{30}}{4} \sqrt{\xi}
 \hskip 1.15cm {\rm for}\ \xi \gg 1, \alpha=6 \nonumber\\
	& =  \ln 2+   \frac{\pi\sqrt{6}}{4} \sqrt{\xi} \hskip 1.8cm {\rm for}\ \xi\gg 1, \alpha=0.
	\label{eq:actionnonminimal_UV}
\end{align} 
The logarithmic contribution of $\xi$ can be naturally parameterized by the perturbative cut-off $\Lambda_{\rm eff}[\xi]$, whose dependence on the value of $\xi$ is given as 
\begin{align}
\Lambda_{\rm eff}[\xi] &= M_P \hskip 1.25cm  {\rm for}\  \xi\ll 1  \nonumber\\ 
 & = \frac{M_P}{\xi} \hskip 1.15cm {\rm for}\ \xi \gg 1, \alpha=6  \nonumber\\ 
 & =\frac{M_P}{\sqrt{\xi}} \hskip 1.15cm {\rm for}\ \xi\gg 1, \alpha=0.
\end{align}
It behaves as  $\Lambda_{\rm eff}[\xi] \sim M_P/\xi$ in the metric formalism, while  $ \sim M_P/\sqrt{\xi}$ in the Palatini formalism. They coincide with the perturbative cut-off scales of the model around the vacuum in each formalism~\cite{Barbon:2009ya, Hertzberg:2010dc, Bezrukov:2010jz, Bauer:2010jg, Antoniadis:2021axu}.

\begin{figure}[t]
	\centering
	\includegraphics[width=.48\textwidth]{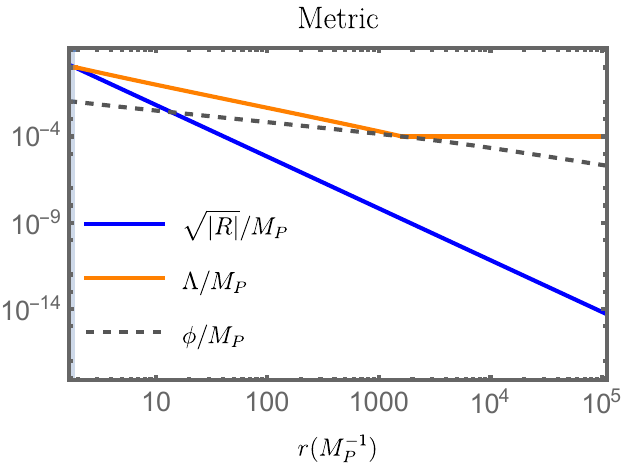} \includegraphics[width=.48\textwidth]{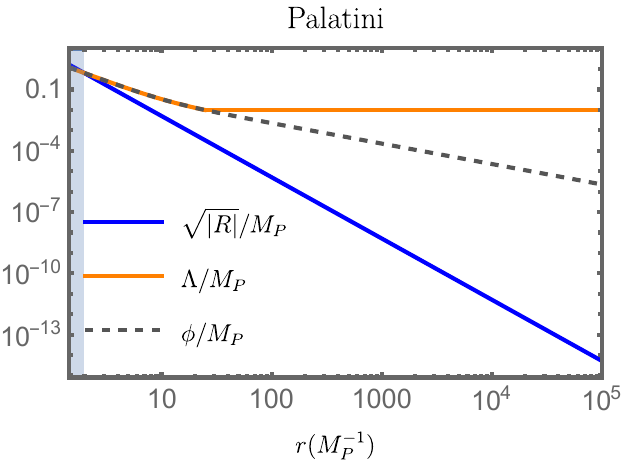} % requires the graphicx package
	\caption{Comparison between the field dependent cutoff $\Lambda(\phi)$ in Eqs.~(\ref{eq:cut_off_metric}), (\ref{eq:cut_off_Palatini}) and the effective energy scale of the wormhole $\sqrt{|R|}$, where $R$ is the scalar curvature Eq.~(\ref{eq:Ricci_scalar}), at a given position $r \in [L_0 , \infty)$  for both metric (left) and Palatini (right) formalisms. The non-minimal coupling is taken as $\xi=10^4$. \dyc{The blue shaded region represents where $\sqrt{|R|}> \Lambda $.}} 
	\label{fig:cutoffcompare}
\end{figure}
%/

We note that, for a large non-minimal coupling, the perturbative cut-off scale is sensitive to the background field value of $\phi$, especially when the axions originate from the underlying $U(1)$ symmetries (not a non-abelian symmetry)~\cite{Bezrukov:2010jz, Antoniadis:2021axu}. We have  
\begin{align} \label{eq:cut_off_metric}
    \Lambda(\phi) &= \frac{M_P}{\xi} \quad {\rm for}\ \phi\ll \frac{M_P}{\xi},\nonumber\\ &= \frac{\xi \phi^2}{M_P}\quad {\rm for}\ \frac{M_P}{\xi}\ll \phi \ll\frac{M_P}{\sqrt{\xi}},\nonumber\\ 
     &=M_P \quad \hskip 0.1cm {\rm for}\  \frac{M_P}{\sqrt{\xi}} \ll \phi
\end{align}
for the metric formalism ($\alpha=6$), while 
\begin{align} \label{eq:cut_off_Palatini}
    \Lambda(\phi) &= \frac{M_P}{\sqrt{\xi}} \quad {\rm for}\ \phi\ll \frac{M_P}{\sqrt{\xi}},\nonumber\\ &= \phi\quad \hskip 0.45cm {\rm for}\ \frac{M_P}{\sqrt{\xi}}\ll \phi \ll M_P,\nonumber\\ 
     &=M_P \quad  \hskip 0.07cm {\rm for}\  M_P\ll \phi
\end{align}
for the Palatini formalism ($\alpha=0$) in the Einstein frame. Therefore, one needs to estimate the validity of the solution by comparing the field dependent cut-off scale $\Lambda(\phi(r))$ and a typical energy scale represented by the scalar curvature $\sqrt{|R(r)|}$  in the range of $r=[L_0, \infty)$.

{ In Fig.~\ref{fig:cutoffcompare}, we compare the two scales in the large $\xi$ limit, along with the profile of the field $\phi(r)$. The $\phi$ field profile has a universal $r$ dependence $\propto 1/r$ away from the throat, then differs for regions closer to the wormhole throat depending on the gravity formalism.  We identify that the energy scale from the curvature $\mu_R\equiv \sqrt{|R|}$ always stays smaller than the local cut-off scale $\Lambda(\phi)$ out of the throat. 
At the throat, $r\sim L_0$, the two scales are about the same~\cite{Abbott:1989jw, Kallosh:1995hi}, which may imply that various quantum corrections
become important. However, the higher-loop corrections are suppressed by the numerical loop-factors, $\sim 1/8\pi^2\sim 10^{-2}$. We leave a systematic study of these corrections near the wormhole throat to future work.

We also note the spacetime regime where $\phi\sim \Lambda(\phi)$. This does not lead to the perturbativity break-down, in general, as long as the higher dimensional terms $\sim C_{4+n}(\phi/\Lambda(\phi))^{n}\phi^4$ are negligible with $C_{4+n}\ll 1$ for all $n\geq 1$.  
In our setup, as given by Appendix~\ref{App:potential} explicitly, the higher dimensional terms in the scalar potential are indeed suppressed by the power of Planck scale $M_P$ in the Jordan frame and further suppressed by the power of $\xi$ in the Einstein frame. Therefore, the perturbative corrections in a regime of background field $\phi\sim \Lambda(\phi)$ for $r\gg L_0$ would not be dangerous.
}

The combination of $\ln \Lambda_{\rm eff}[\xi]/\phi$ in the wormhole action Eq.~(\ref{eq:actionnonminimal}) can provide a more closed expression of the $U(1)_{\rm PQ}$ breaking non-perturbative factor as 
\begin{align}
	e^{-S_{wh} + in\theta } &= \exp\left[- 
 n \left(S_{wh}^{\rm UV}[\xi] +  \ln\frac{\Lambda_{\rm eff}[\xi]}{\phi} - i \theta \right)\right]  = \left( \sqrt{2}\, e^{- S_{wh}^{\rm UV}[\xi]}  \frac{\Phi}{\Lambda_{\rm eff}[\xi]}\right)^n . 
  \label{eq:holomorphic_correction}
\end{align}
This combination will enter the effective Lagrangian of $\Phi$. At this moment,
the overall scale of the prefactor $M_*^4$ that will be multiplied by the local operator Eq.~(\ref{eq:holomorphic_correction}) is not trivial. 
Conservatively we can think of it as just ${\cal O}(M_P^4)$, but it can change greatly depending on the cut-off scales in the vacuum like ${\cal O}(L_0^{-4}, \Lambda^4_{\rm eff}[\xi])$. We leave the detailed estimation to future work.

The form of Eq.~(\ref{eq:holomorphic_correction}) clearly shows that 
(i) the contribution of a wormhole with a PQ charge of $n$ is equivalent to $n$ times the contribution of a unit-charged wormhole, (ii) 
when combined with an axion-dependent term, the primary contribution is the holomorphic function of $\Phi$. Each properties are not independent of each other. 
In our example, the UV contribution is nearly independent of the {\bf IR} field value, so the IR contribution is quite similar to the case of an extremal instanton with $L_0=0$. Since $f^2_a(\phi)\ll f^2_a(\phi_0)$ with $\phi\ll \phi_0$, we have $p(\phi)\simeq -n/f_a(\phi)$ and  
\begin{align} 
S^{\rm IR}_{wh}[n,\phi] \simeq \int^\phi_{\phi_0}  d\varphi\, p(\varphi)   \simeq  n\int_\phi^{\phi_0}  \frac{d\varphi}{f_a(\varphi)}.
\end{align}
The holomorphicity of the local operator is quite general even for multiple axion cases if the UV contribution of the wormhole (contribution near the throat) is nearly independent of the {\bf IR} field values for the radial scalar partners.

Before closing this section, let us briefly discuss different types of axion models. When an axion comes from string theory~\cite{Svrcek:2006yi}, the radial scalar partner usually has a geometrical meaning. In that case, for $G(\phi)=1$ in Eq.~(\ref{eq:single_axion_model}), the axion decay constant has the form $f_a(\phi) = f_a e^{\beta \phi/M_P}$ with $\beta={\cal O}(1)$, a dilaton-like coupling. 
It is well known from many studies~\cite{Giddings:1987cg, Kallosh:1995hi, Heidenreich:2015nta, Hebecker:2016dsw, Brax:2023now} that the corresponding axion wormhole action depends less on the dynamics of the scalar partner than in the case we discussed. 
For the change of the radial field value $\Delta \phi \sim M_P$ in the wormhole background, the axion decay constant does not change significantly, i.e. $\Delta f_a(\phi)/f_a(\phi) ={\cal O}(1)$. Therefore, the parameter dependence of the action is more similar to the Giddings-Strominger wormhole action, $S_{wh}\sim M_P/f_a$. Since the throat size is sensitive to the IR field value, 
the holomorphicity of the wormhole induced operator does not hold in general.  
We can first define  a complex scalar field 
\begin{align} 
T(x)= S(x)-i \theta(x)
\end{align}
to represent the kinetic term of the axion and its radial partner as
\begin{align}
    \frac{1}{2}(\partial_\mu\phi)^2 + \frac{1}{2} f_a^2(\phi) (\partial_\mu\theta)^2 
    =  \frac{2M_P^2}{\beta^2 (T+T^*)^2} (\partial_\mu T)(\partial^\mu T^*).
\end{align}
It can be easily proved that the wormhole solution exists only for $|\beta| < \sqrt{2/3}$.
From Eqs.~(\ref{eq:boundary condition_single}), (\ref{eq:wormhole_action_single_axion}) we have
\begin{align} 
S_{wh} = \frac{n M_P \sin  (\frac{\pi\sqrt{6}}{4}\beta )}{\beta f_a(\phi)}
= n  \sin \left(\frac{\pi\sqrt{6}}{4}\beta\right) S, 
\end{align}
where $S={\rm Re}(T)$. 
This gives 
\begin{align}
    e^{-S_{wh} + in\theta}= e^{- nT} \times \exp\left[ \frac{n}{2} 
    \Big( 1- \sin\Big(\frac{\pi\sqrt{6}}{4}\beta\Big)\Big) (T+T^*) \right].
\end{align}
The non-holomorphic piece that exists in addition to the holomorphic part $e^{-nT}$ disappears when $\beta =\sqrt{2/3}$, i.e. $L_0=0$. In this sense, the fact that many examples for axions in string theory give $\beta \geq \sqrt{2/3}$ could be related to the condition of preserving supersymmetry  

\subsection{Complex scalar model with an additional scalar field} 
Apart from a complex scalar $\Phi$, other real scalar fields $\chi$ that couple to $\Phi$ can exist. One of the well-motivated forms of the coupling is that 
the additional scalar couples to the kinetic term of the complex scalar as a product form $Z(\chi)\times K_{\rm kin}(\Phi, \Phi^*)$. 

In this section, we  discuss how such an additional scalar field changes the wormhole action. 
Our starting action is given by
\begin{align} \label{eq:twoscalar_model}
	\mathcal{S} &= \int d^4 x \sqrt{-g} \left( - \frac{M_P^2}{2} R  
	+ \frac{1}{2} Z(\chi) K_{\rm kin}(\Phi, \Phi^*) +\frac{1}{2}   \partial_\mu\chi \partial^\mu\chi  \right) \nonumber\\
	 &= \int d^4 x \sqrt{-g} \left[ - \frac{M_P^2}{2} R  
+  \frac{1}{2} Z(\chi)\Big(   \partial_{\mu} \phi \partial^{\mu} \phi  +   f^2(\phi)  \partial_{\mu} \theta  \partial^\mu \theta \Big) +\frac{1}{2}   \partial_\mu\chi \partial^\mu\chi  \right] .
\end{align}
For a given complex scalar, $\phi$ is the radial partner of the axion $\theta$. 
$\phi$ and $\chi$ are properly normalized without introducing a kinetic mixing between them.
Considering Eq.~(\ref{eq:general action}), we identify fields as $\phi^1=\phi$, $\phi^2=\chi$, $\theta^1=\theta$, and the metric of the kinetic term as $G_{11}=Z(\chi)$, $G_{12}=G_{21}=0$, $G_{22}=1$, $(f^2_a)_{11}= Z(\chi) f^2(\phi)$.

From the equations of motion for the Euclidean action, 
we have the relation between the throat radius and the throat field values as 
\begin{align}\label{eq:additional scalar throat }
    L_0^2  =  \frac{1}{2\pi^2 \sqrt{6}}  \left(\frac{n}{ M_P \sqrt{Z(\chi_0)}f(\phi_0)}\right).
\end{align}
From Eqs.~(\ref{eq:eom_for_phi1}), (\ref{eq:eom_for_phi2}),
the equations of motion for $\phi$ and $\chi$ are 
\begin{align} \label{eq:eom for phichi1}
&  \left(\frac{d\chi}{d\tau}\right)^2 + Z(\chi)\left(\frac{d\phi}{d\tau}\right)^2  =	\frac{n^2}{Z(\chi)f^2(\phi)} -  \frac{n^2}{Z(\chi_0) f^2(\phi_0)}, \nonumber\\
 &\hskip 0.2cm	\frac{d^2 \chi}{d\tau^2} -\frac{1}{2} \frac{dZ}{d\chi} \left(\frac{d\phi}{d\tau}\right)^2 +\frac{1}{2} \frac{dZ}{d\chi}\left(\frac{n^2}{Z^2(\chi)f^2(\phi)}\right)  =0.
\end{align}
With the boundary conditions $\phi'(0)=\chi'(0)=0$, these equations are simplified as
\begin{align} \label{eq:sol of chi}
  \left(\frac{d\phi}{d\tau}\right)^2& =	\frac{n^2}{Z^2(\chi)}\left(\frac{1}{f^2(\phi)} - \frac{1}{f^2(\phi_0)}\right),\nonumber\\
     \left(\frac{d\chi}{d\tau}\right)^2 &= \frac{n^2}{f^2(\phi_0)}\left(\frac{1}{Z(\chi)} -\frac{1}{Z(\chi_0)}\right).
\end{align}
For given \textbf{IR} fields $\chi$ and $\phi$,  the throat field values $\phi_0$ and $\chi_0$ 
are determined by  
\begin{align}  \label{eq:boundary condition_phichi}
	\int_\phi^{\phi_0} d\phi \frac{1}{\sqrt{f^2(\phi_0)/f^2(\varphi)-1}}  &= \int_\chi^{\chi_0} \frac{dx}{\sqrt{Z(x)}} \frac{1}{\sqrt{1-  Z(x)/Z(\chi_0)}},\nonumber\\
	\int_\chi^{\chi_0}  \frac{dx}{M_P} \frac{1}{\sqrt{Z(\chi_0)/Z(x) - 1}} &= \frac{\pi\sqrt{6}}{4},
\end{align} 
assuming $\phi_0>\phi$ and $\chi_0>\chi$.
With the solutions, the wormhole action becomes
\begin{align}
	S_{wh}[n,\phi,\chi] = \int_0^{\tau_\infty} d\tau \frac{n^2}{Z(x)f^2(\varphi)} 
	= n \int_\phi^{\phi_0} \frac{d\varphi}{f(\varphi)}\frac{1}{\sqrt{1- f^2(\varphi)/f^2(\phi_0)}}.
\end{align}
Notice that the general expression for the wormhole action does not explicitly depend on $Z(\chi)$. 
However, the existence of $\chi$ affects the value of $\phi_0$, so the form of $S_{wh}$ can have a nontrivial dependence on $\chi$.
Being equipped with the general formulation for this case, we now present some specific examples and examine their implications. 

\subsubsection{Model with a light dilaton \label{subsubsection_complex_scalar_and_dilaton}}

Let us consider a dilatonic coupling for the complex scalar field $\Phi=\frac{1}{\sqrt{2}}\phi e^{i\theta}$~\cite{Alvey:2020nyh}, so 
\begin{align} 
	Z(\chi) = e^{2\beta\chi/M_P},\quad f(\phi) = \phi.
\end{align}
Here we take a convention $\beta\geq 0$.
Using Eq.~(\ref{eq:boundary condition_phichi}), we get 
\begin{align}
\exp\left(-\frac{\beta(\chi_0 -\chi)}{M_P} \right) = \cos\left(\frac{\pi\sqrt{6}}{4}\beta\right).
\end{align}
The wormhole solution is only available for $\beta < \sqrt{2/3}$. We also obtain 
\dyc{\begin{align}
	 \frac{\phi_0}{M_P} \sqrt{1- \frac{\phi^2}{\phi_0^2}} =  \frac{e^{-\beta \chi / M_P}}{\beta}  \sin\left(\frac{\pi\sqrt{6}}{4}\beta\right) = \frac{e^{-\beta \chi_0 / M_P}}{\beta} \tan\left(\frac{\pi\sqrt{6}}{4}\beta\right) .
\end{align}}
The wormhole throat radius becomes 
\begin{align}
	L_0^2  & = \frac{1}{2\pi^2\sqrt{6}}\left(\frac{n}{ M_P e^{\beta\chi_0} \phi_0}\right) = \frac{1}{2\pi^2\sqrt{6}}\left(\frac{n }{M_P^2 \tan(\frac{\pi\sqrt{6}}{4}\beta)/\beta}\right).
\end{align}
It is confirmed that the wormhole throat size is small $L_0 \lesssim 1/M_P$, so the UV contribution is just $O(1)$. For $\phi \ll \phi_0$, the corresponding wormhole action can be represented by the function of $\phi$ and $\chi$ as
\begin{align} \label{eq:wormhole_action_two_scalars}
   S_{wh}[n,\phi,\chi] & = n \int_\phi^{\phi_0} \frac{d\phi}{\phi} \frac{1}{\sqrt{1- \phi^2/\phi_0^2 }} 
  \simeq   n \ln \left(\frac{2 M_P \sin(\frac{\pi\sqrt{6}}{4} \beta )/\beta }{ e^{\beta\chi} \phi }\right)  .
\end{align}
The IR dominated wormhole action has the form of $S_{wh}\simeq n \ln M_P/f_a(\chi,\phi)$, 
where the effective axion decay constant is provided by $f_a(\phi, \chi) = e^{\beta\chi}\phi$. 
Therefore, the overall value is log-enhanced, almost always leading to an at most $\mathcal{O}(10)$ value. 
\dyc{This indicates that this case ($n=1$) lies beyond the regime of validity of the effective theory. Only a large wormhole solution with a large enough $n$ with the radius size greater than ${\cal O}(10)$ can be trusted.}
This particular case can also be interpreted in the sense that the dilaton field $\chi$ alters the axion decay constant to an effective value $f_{a}$.
The PQ breaking local operator is proportional to 
\begin{align}
	e^{-S_{wh} + in\theta } &=  \left(  \frac{e^{\beta\chi}\Phi}{\sqrt{2}M_P \sin(\frac{\pi\sqrt{6}}{4}\beta)/\beta}\right)^n . 
	\label{eq:holomorphic_correction2}
\end{align} 
 
\subsubsection{Model with a $R^2$-term}
The previous type of action can naturally arise from a theory consisting of a complex scalar non-minimally coupled to gravity, along with a $R^2$-term\footnote{This class of theories are also considered as a UV extension to theories incorporating non-minimal couplings to gravity~\cite{Ema:2017rqn, Gorbunov:2018llf, He:2018gyf, He:2018mgb}. }:
\begin{align} \label{eq:nonminimal_scalaron}
	\hskip -0.1cm \mathcal{S} = \int d^{4} x \sqrt{-g} \left[-\left(\frac{M_P^2}{2 } +  \xi_{\phi}|\Phi|^2\right)R + \frac{\xi_s}{4}R^2 + \partial_\mu\Phi \partial^\mu\Phi^*
  - V(|\Phi|) \right]. 
\end{align}
In metric formalism, introducing the $R^2$ term leads to a new degree of freedom in the Einstein frame. 
The equivalent form of the action becomes 
\begin{align}
  \hskip -0.15cm  {S} =& \int d^4x \sqrt{-g}\left[-\frac{M_P^2}{2} R  + \frac{1}{2} e^{\sqrt{\frac{2}{3}} \frac{\chi}{M_P}}  \left(\partial_{\mu } \phi \partial^{\mu}\phi + \phi^2 \partial_{\mu}\theta \partial^{\mu} \theta  \right)   \right. \nonumber\\ 
  &    \hskip 2.1cm \left. +\, \frac{1}{2}  \partial_{\mu} \chi \partial^{\mu} \chi - U(\chi, \phi) \right]
\end{align}
with the potential term containing the $\chi, \phi$ dependence as the form
\begin{align}
    U(\chi, \phi) =  \frac{M_P^2}{4\xi_s}\left[1 - e^{\sqrt{\frac{2}{3}}\frac{\chi}{M_P}}\left(1+\frac{\xi_{\phi}\phi^2}{M_P^2} \right) \right]^2 + e^{2\sqrt{\frac{2}{3}}\frac{\chi}{M_P}}V(\phi) .
\end{align}
For the following discussion, we neglect the effect of $V(\phi)$ as we did in the previous sections. However,
the first term of $U(\chi, \phi)$ cannot be simply ignored. 
Around the minimum, the mass of $\chi$ is $m_s= {\cal O}(M_P/\sqrt{\xi_s})$. 
If the mass of $\chi$ is smaller than $M_P/\xi_\phi$,  $\chi$ plays a role of unitarizing the action up to the Planck scale at the vacuum.  

\begin{figure}[htbp]
   \centering
   \includegraphics[width=.65\textwidth]{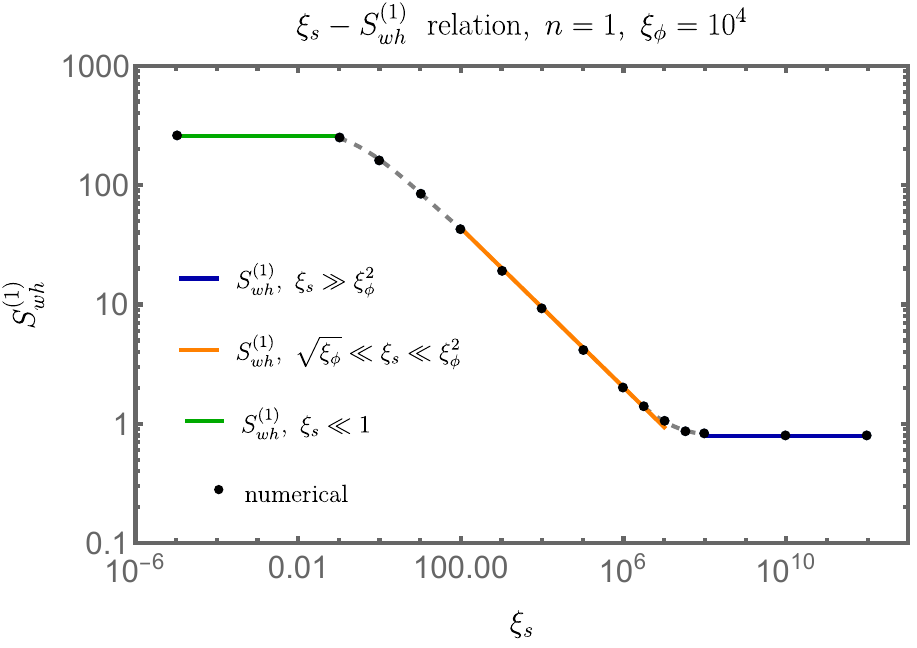} % requires the graphicx package
\caption{$S_{wh}^{(1)}$ for $\xi_\phi = 10^{4}$. The black dots represent the numerical computation, the orange line corresponds to Eq.~(\ref{eq:actionR2}), the green line corresponds to the value when $\xi_s \ll 1$, and the blue line corresponds to the value obtained with Eq.~(\ref{eq:additional scalar throat }). We note a transition around $\xi_s \simeq \xi_\phi^2$, and another transition expected around $\xi_s \lesssim \sqrt{\xi_{\phi}}$, both depicted in dashed - gray lines. }
   \label{fig:dilatonR2action}
\end{figure}

Let us consider cases of large $\xi_\phi$. 
When $\xi_s\to 0$, $\chi$ is supermassive, therefore it is just frozen with the condition $\partial_\chi U(\chi, \phi)=0$.  In this limit, we recover the single complex scalar model, Eq.~(\ref{eq:complexscalar_model}). Thus, $S_{wh} \sim \sqrt{\xi_\phi} + \ln M_P/(\xi_\phi\phi)$.
On the other hand, in the opposite limit  $\xi_s\to \infty$, $\chi$ is effectively massless. Hence, the action becomes the form of Eq.~(\ref{eq:twoscalar_model}).
In this case,  even for a large value of $\xi_\phi$, $S_{wh} \sim \ln M_P/(e^{\beta\chi/M_P}\phi)$ with $\beta=1/\sqrt{6}$.
This implies that there is a screening effect of $\xi_\phi$
as $\xi_s$ increases. More specifically, we notice that in the regime where $\xi_\phi \phi^2 \gtrsim M_P^2$, $U(\chi, \phi)$ resembles a quartic potential 
\begin{align}\label{eq:potentialapproxR2}
    U(\chi, \phi) \simeq\frac{1}{4} e^{2\sqrt{\frac{2}{3}}\frac{\chi}{M_P}} \left(\xi_\phi^2/\xi_s\right)  \phi^4 .
\end{align}
From the {\bf IR} value of $\chi$, the displacement of $\chi$ in the wormhole background is at most of ${\cal O}(M_P)$, so that $ e^{2\sqrt{\frac{2}{3}} \frac{\chi}{M_P}} = {\cal O}(1)$. 
When the quartic coupling $\lambda =\xi_\phi^2/\xi_s\ll 1$, this potential term is irrelevant and the system reduces to the previous example in Sec.~\ref{subsubsection_complex_scalar_and_dilaton}. 
It results in a small wormhole action as Eq.~(\ref{eq:wormhole_action_two_scalars}). As $\lambda$ gradually increases ($\xi_s$ decreases), the potential is approximated by a quartic term of $\phi$ with a very large self coupling. In this case, the throat size  explicitly depends on the values of $\xi_s$ and $\xi_\phi$~\cite{Abbott:1989jw}, giving the form
\begin{align}
    L_0^2 \simeq  \frac{n}{8\pi^2 M_P^2} \left(\frac{n\, \xi_\phi^2}{2\pi^2\xi_s}\right)^{1/3}   \quad{\rm for} \ \sqrt{\xi_\phi} \ll \xi_s \ll \xi_\phi^2 .
\end{align}
In turn, this leads to the throat part of the wormhole action 
\begin{align}
    S_{wh}^{(1)} = 3\pi^3 M_P^2 L_0^2 \simeq \frac{3n}{8 }  \left(\frac{n\pi  \xi_\phi^2}{2 \xi_s}\right)^{1/3}
    \quad{\rm for} \ \sqrt{\xi_\phi} \ll \xi_s \ll \xi_\phi^2 .
    \label{eq:actionR2}
\end{align}
This $S_{wh}^{(1)}$ is depicted in Fig.~\ref{fig:dilatonR2action}. 
For a large value of $\xi_\phi$, 
the analytic form of $S_{wh}^{(1)}$ are different in three limiting cases, 
(I) $\xi_s \ll \sqrt{\xi_\phi} $ (Green), (II) $ \sqrt{\xi_\phi} \ll \xi_s \ll \xi_\phi^2 $ (Orange), (III) $\xi_\phi^2\ll \xi_s$ (Blue).

Therefore, the value of $\xi_s$ is quite important to determine the size of axion wormhole action when $\xi_\phi \gg 1$. Actually, in the metric formalism, 
 the natural value of $\xi_s$  is proportional to $\xi_\phi^2$ because of the quantum corrections from the following renormalization group equation at one-loop level~\cite{Elizalde:1993ee, Elizalde:1993ew, Salvio:2014soa, Salvio:2015kka, Codello:2015mba, Markkanen:2018bfx, Ema:2019fdd, Ema:2020evi}. 
\begin{align}
	\mu\frac{d\xi_s}{d\mu} = 
	-\frac{1}{4\pi^2}\left(\xi_\phi +\frac{1}{6}\right)^2.
\end{align}
This implies that even if the tree-level value of $\xi_s$ is vanishing, we expect $\xi_s$ is naturally generated as $\xi_s\sim 0.02\, \xi_\phi^2 \ln\Lambda/\mu \sim 0.02 \xi_\phi^2$. Inserting this to the wormhole action, Eq.~(\ref{eq:actionR2}), we have 
$S_{wh}^{(1)} \simeq 1.5$ ($n=1$). 
This shows the difficulty of solving the axion quality problem with the help of the non-minimal coupling to gravity for the complex scalar field that includes the axion as $\Phi= \frac{1}{\sqrt{2}} \phi e^{i\theta}$.

Before closing this section, we briefly comment on the case of Palatini formalism. 
In the Palatini formalism, there is no new scalar degree of freedom even if we introduce $R^2$-term (see~\cite{Sotiriou:2008rp} for a review).
In the Einstein frame, Eq.~(\ref{eq:nonminimal_scalaron}) becomes 
\begin{align}
	S&=\int d^4 x \sqrt{-g}\left( - \frac{M_P^2}{2} R 
	+ \frac{ |\partial_\mu\Phi|^2 
	}{(1+\xi_\phi\phi^2/M_P^2)}   + \frac{ \xi_s |\partial_\mu\Phi|^4}{2(1+\xi_\phi\phi^2/M_P^2)^2M_P^4}   \right).
\end{align}  
Because of nontrivial higher derivative terms, 
it is more difficult to figure out the effect of large $\xi_s$. 
From a naive dimensional analysis, $\partial \phi \sim \phi/r$, 
and for $\xi_\phi \phi^2 \gtrsim M_P^2$, 
the higher derivative term can be important when 
the distance becomes scale as 
$r \lesssim \sqrt{\xi_s/\xi_\phi} M_P^{-1}$.
By comparing it with the throat radius  in the small $\xi_s$ limit $L_0\sim \xi_\phi^{1/4} M_P^{-1}$, 
we see that the higher derivative terms are not effective in the regime $L_0 \leq r< \infty$ for $\xi_s^2 \ll \xi_\phi^3$, while it can provide a nontrivial effect in the opposite limit.  We leave the study for the Palatini formalism to a future work.

\section{Discussion}

In this study, we employ the effective theory approach with a complex scalar field, which incorporates the axion  as its complex phase degree,  
 to investigate the Euclidean wormhole action, where the stability of the wormhole throat relies on the presence of a non-zero Peccei-Quinn charge associated with axions. Our analysis yielded a clear understanding of both the parameter dependence of the UV contribution and the IR behavior of the wormhole action. We achieve this understanding by explicitly calculating the dependency of the infrared ({\bf IR}) degrees of freedom in axion models featuring radial scalar partners.
In particular, when a significant non-minimal coupling of the scalar field with gravity is present, our analysis reveals that the UV contribution of the wormhole action becomes substantial, reaching the order of $S_{wh}^{\rm UV} = \mathcal{O}(\sqrt{\xi})$. This substantial wormhole action, surpassing a value of approximately $\mathcal{O}(200)$ with $\xi\gtrsim  10^4$, offers a straightforward solution to the axion quality problem.

The reason behind the substantial magnitude of the action in the case of a large non-minimal coupling lies in the fact that, unlike the minimally coupled model with $\xi=0$, the dynamics of the radial field becomes independent of that of the axion as the radial field surpasses a critical threshold. Consequently, as we approach the wormhole throat from the asymptotic region, the axion's decay constant simply converges to $M_P/\sqrt{\xi}$. This decoupling phenomenon between the radial scalar field and the axion field appears to be rooted in symmetry considerations and may also be associated with the conditions of the scalar potential required for inflation. However, we demonstrated that the effects of a large non-minimal coupling can be  mitigated by introducing the $R^2$-term with a specific range of coupling values. The true impact of a large $\xi$ warrants a more detailed investigation in both the metric and Palatini formalisms.

If additional gauge symmetries are present, the QCD axion must necessarily be a gauge singlet, formed as a combination of pseudo-Nambu-Goldstone bosons, denoted as $\theta(x) = \sum_I q_I \theta^I(x)$. Other combinations are absorbed by the gauge bosons. For complex scalar fields $\Phi_I = \frac{1}{\sqrt{2}}\phi_I e^{i\theta^I}$, which carry charges under these gauge symmetries, this results in local operators proportional to $e^{-S_{wh} + i q_I\theta^I} \simeq e^{-S_{wh}^{\rm UV}}\prod_I (\Phi_I/\Lambda_{\rm eff})^{q_I}$.
On one hand, if the axion is a composite particle, evaluating the wormhole solution becomes highly intricate. This complexity arises because we must account for strongly interacting fermions in the equations of motion.
Subsequent research will provide a deeper connection between more fundamental theory and axion phenomenology.

\acknowledgments

We thank Miguel Escudero, Koichi Hamaguchi, Hyun Min Lee, Natsumi Nagata, and Hyeonseok Seong for valuable discussions. 
SCP is supported by National Research Foundation grants funded by the Korean government (NRF-2021R1A4A2001897) and (NRF-2019R1A2C1089334).
CSS acknowledges support from (NRF-2022R1C1C1011840, NRF-2022R1A4A5030362) and IBS under the project code, IBS-R018-D1. DYC also acknowledges support from the CERN-Korea Theoretical Physics Collaboration and Developing Young High-Energy Theorists fellowship program (NRF-2012K1A3A2A0105178151).

\appendix

\section{The effect of the scalar potential on the axion wormhole action} \label{App:potential}

In this appendix, we discuss the effect of the scalar potential including higher dimensional operators on the axion wormhole solution in more detail. Throughout the appendix, instead of the singular coordinate for the wormhole geometry, we take the regular one as
\be
ds^2_{wh} = \frac{dr^2}{(1- L_0^4/r^4)} + r^2 d\Omega_3^2 = d\rho^2 + a^2(\rho) d\Omega_3^2
\ee
where
\be
\frac{d\rho}{dr}= \frac{1}{\sqrt{1- L_0^4/r^4}},\quad a(\rho)=r .
\ee
 At the wormhole throat ($\rho =0$), $a(\rho)= L_0$, and $a(\rho)=r \simeq \rho$ in the area far away from the wormhole ($\rho\gg L_0$).

Considering a complex scalar field that contains an axion $\theta(x)$, 
\be
\Phi=\frac{1}{\sqrt{2}} \phi\, e^{ i\theta},
\ee 
we can add a potential term $V_J(|\Phi|)$ to the action Eq.~(\ref{eq:complexscalar_model})  as
\be	 \label{eq:action_with_pot}
\mathcal{S} = \int d^4 x \sqrt{-g} \left[ - \left(\frac{M_P^2}{2} + \xi |\Phi|^2\right) R +     \partial_{\mu} \Phi \partial^{\mu} \Phi^*  - V_J(|\Phi|) \right] .
\ee
In the Einstein frame, the Euclidean action for the matter part
is given as 
\be
(\mathcal{S}_E)_{\rm mat} = 2\pi^2 \int_0^\infty d\rho\, a^3 \left(\frac{1}{2} G(\phi) \left(\frac{d\phi}{d\rho}\right)^2 +  
\frac{n^2}{8\pi^4 f_a^2(\phi) a^6}  + V(\phi)  \right), 
\ee
where $G(\phi)$ and $f_a(\phi)$ are given by Eq.~(\ref{eq:kinetic_metric}). 
Here the potential in the Einstein frame is given by $V(\phi)= V_J(\phi)/(1+\xi\phi^2/M_P^2)^2$. The PQ charge quantization 
of the wormhole (Eq.~(\ref{eq:PQcharge})) is used for the axion kinetic term as 
\be
2\pi^2 f_a^2(\phi) a^3 \left(\frac{d\theta_E}{d\rho}\right) = n,
\ee
where $\theta_E$ is the Euclidean dual of the $3$-from field strength $H_{\mu\nu\rho}$. 
Then, the equations of motion for $\phi$ 
are 
\be
G(\phi)\left(\frac{d^2\phi}{d\rho^2} + \frac{3}{a}\frac{da}{d\rho} \frac{d\phi}{d\rho}\right) 
+\frac{1}{2}\frac{dG(\phi)}{d\phi} \left(\frac{d\phi}{d\rho}\right)^2 + \frac{n^2}{4\pi^4 f_a^3(\phi)a^6}\frac{df_a(\phi)}{d\phi}
= \frac{dV(\phi)}{d\phi}.
\ee
The Einstein equation gives 
\be\label{app:Eeq}
\left(\frac{da}{d\rho}\right)^2 
 = 1-\frac{a^2}{3 M_P^2}  
\left( \frac{n^2}{8\pi^4 f_a^2(\phi) a^6}
- \frac{1}{2}G(\phi)\left(\frac{d\phi}{d\rho}\right)^2 + V(\phi)\right).
\ee
As $\rho\to 0$, $a(\rho)\to L_0$, the derivatives  $da/d\rho,\, d\phi/d\rho\to 0$. 
From the solutions of these equations, the contribution to the wormhole action can be written as 
\be
S_{wh} = \int_0^\infty d\rho \left(\frac{dS_{wh} }{d\rho}\right).
\ee
The potential contribution to the wormhole action can be separated as 
\be
\Delta S_{wh} =  \int_0^\infty d\rho \left(2\pi^2 a^3 V(\phi)\right). 
\ee
For the scalar potential in the Jordan frame, $V_J$ is taken as the following form
\be \label{eq:potential_Jordan}
V_J(\phi)=  \frac{\lambda}{4}\Big(\phi^2 - f_a^2 \Big)^2
\Big( 1 + \frac{\lambda_6}{M_P^2}\phi^2 + \cdots  \Big).
\ee
To simplify the numerical calculation, we choose $f_a$ as a vacuum expectation value of $\phi$ for general values of $\lambda$ and $\lambda_6$. We include the higher dimensional term with a natural cut-off scale $M_P$ in the Jordan frame. The perturbative cut-off $\Lambda(\phi)$ in the Einstein frame could differ from $M_P$ depending on the background value of $\phi$. However, it does not mean the cut-off scale in the potential term of Eq.~(\ref{eq:potential_Jordan}) should be directly dependent on $\Lambda(\phi)$.

\begin{figure}[h]
    \centering
    \includegraphics[width=0.6\linewidth]{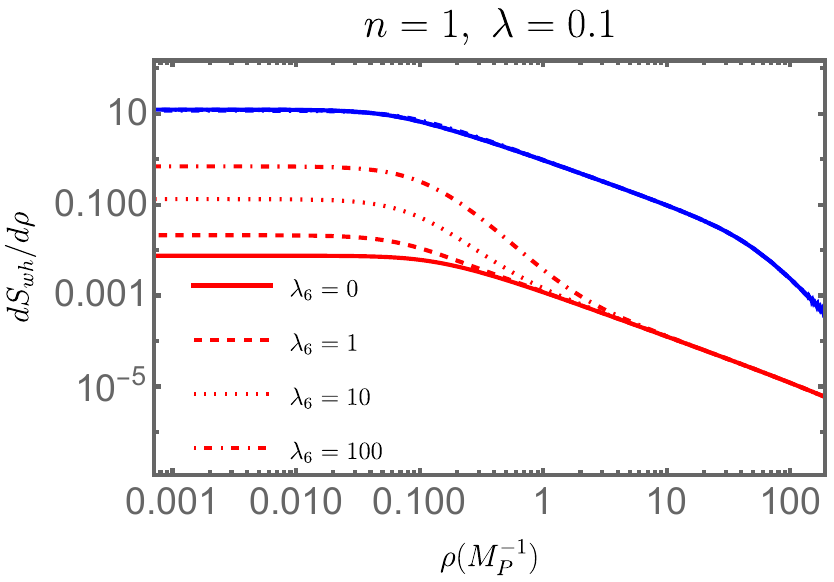}
    \caption{The $dS_{wh}/d\rho$ contribution for minimal gravity, evaluated with $f_a=10^{16}\,{\rm GeV}$, $n=1$, and $\lambda = 0.1$. The blue lines depict the total contribution, whereas the red lines correspond to the contribution from the potential ($2\pi^2 a^3 V(\phi)$), with different values of $\lambda_{6}$. }
    \label{fig:action_contribution_minimal}
\end{figure}

\begin{figure}[h]
    \centering
    \includegraphics[width=0.48\linewidth]{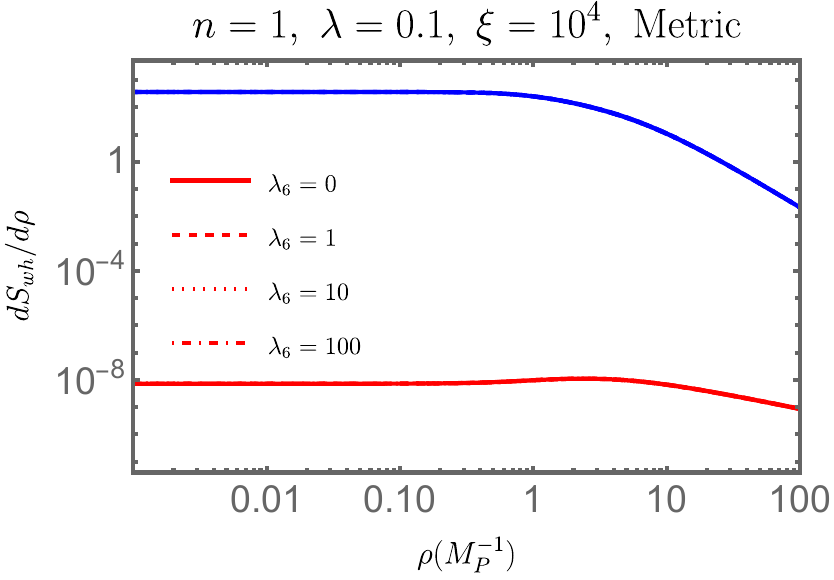}
    \includegraphics[width=0.48\linewidth]{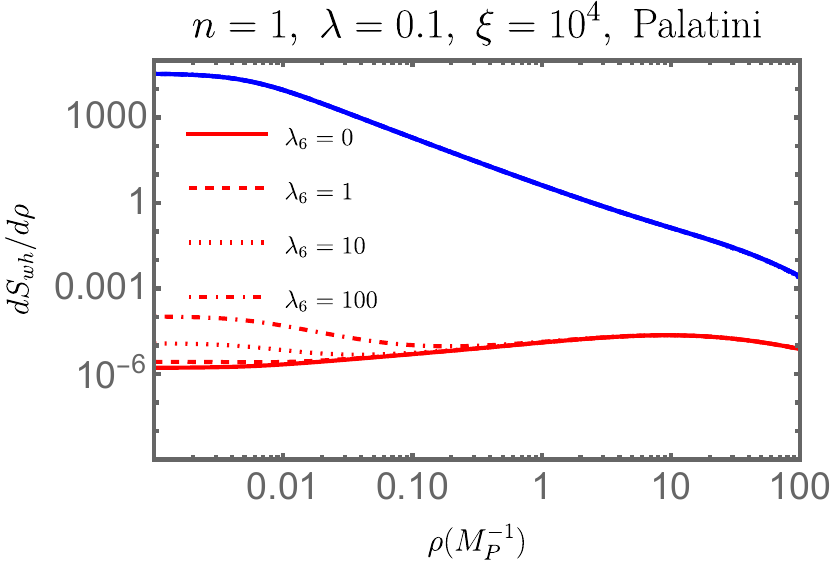}
    \caption{The $dS_{wh}/d\rho$ contribution for non-minimal gravity in the metric formalism(left), and non-minimal gravity in the Palatini formalism(right). All cases were evaluated with $f_a=10^{16}\,{\rm GeV}$, $n=1$, $\lambda = 0.1$, and $\xi = 10^{4}$, using the same color scheme as Fig.~\ref{fig:action_contribution_minimal}. }
    \label{fig:action_contribution_nonminimal}
\end{figure}

In Fig.~\ref{fig:action_contribution_minimal} and Fig.~\ref{fig:action_contribution_nonminimal}, we depict the action contribution $dS_{wh}/d\rho$ including higher-order operators from numerical calculations incorporating a shooting method with the potential Eq.~(\ref{eq:potential_Jordan}) . We see that for all cases, the potential contribution $\Delta( dS_{wh}/ d\rho) = 2\pi^2 a^{3} V(\phi)$ is subdominant compared to the total value by orders of magnitude, which makes the wormhole solution insensitive to the potential form. 
Although they are subdominant, the potential contribution becomes maximal around the wormhole throat because the value of $\phi$ increases as $\rho\to 0$. Therefore, it is sufficient to discuss 
the contribution from the potential term focusing around the wormhole throat.  

From Eq.~(\ref{app:Eeq}), the throat radius $L_0$ is given by 
\be
L_0^2 = \frac{1}{2\pi^2\sqrt{6}}
\left(\frac{n}{M_P f_a(\phi_0)}\right) \left(1-\frac{V(\phi_0) L_0^2}{3M_P^2}\right)^{-1/2}, 
\ee 
where $\phi_0=\phi(\rho)|_{\rho\to 0}$. 
Therefore, as the potential contribution increases, the size of the wormhole also increases. 
Compared to the massless scalar case discussed in the text, the contribution of the potential term is suppressed by the factor
\be \label{eq:pot_cont}
\frac{\Delta S_{wh}}{S_{wh}} =\frac{V(\phi_0) L_0^2}{6M_P^2}\simeq \frac{V(\phi_0)}{12\pi^2\sqrt{6} M_P^3 f_a(\phi_0)}. 
\ee 
One can directly insert the values of $\phi_0$ in Eq.~(\ref{eq:phi0}) for each model to Eq.~(\ref{eq:pot_cont}) and estimate its magnitude. Considering the scalar potential Eq.~(\ref{eq:potential_Jordan}) and $f_a(\phi)$ in Eq.~(\ref{eq:kinetic_metric}), 
we have 
\be
\frac{\Delta S_{wh}}{S_{wh}}
= \frac{1}{12\pi^2 \sqrt{6}}\left(\frac{V_J(\phi_0)}{ M_P^4}\right)\left(\frac{M_P}{(1+\xi \phi_0^2/M_P^2)^{3/2}  \phi_0}\right).
\ee
For each case, i.e. $\xi\ll 1$ and $\xi\gg 1$ 
in the metric ($\alpha=6$) or Palatini ($\alpha=0$)  formalism, 
we have 
\be
\hskip -1cm\frac{\Delta S_{wh}}{S_{wh}} &=& \frac{1}{18\pi^3 M_P^4} 
 V_J\left(\frac{\pi\sqrt{6}}{4} M_P\right)\simeq 0.006\lambda\Big(1+3.7 \lambda_6\Big) 
\hskip 1.8cm (\xi\ll 1) \\ 
&= & \frac{4\sqrt{\xi}}{81\sqrt{30} \pi^2 M_P^4}  V_J\left(\sqrt{\frac{5}{4\xi}} M_P\right)  \simeq \frac{0.0004\lambda}{\xi^{3/2}}\left(1+1.25 \frac{\lambda_6}{\xi}\right)
\hskip 0.35cm (\xi\gg 1, \, \alpha=6) \\
 & =& \frac{1}{27\sqrt{6}\pi^2\xi^{3/2}M_P^4} V_J\left(\sqrt{\frac{3}{2}} M_P\right)\simeq \frac{0.0009 \lambda}{\xi^{3/2}}\Big(1 +1.5\lambda_6\Big)
\hskip 0.6cm (\xi\gg 1, \, \alpha=0)
\ee for $\phi_0 \gg f_a$. 
These corrections depending on the couplings $\lambda$, $\lambda_6$ of the scalar potential and the non-minimal coupling $\xi$ are consistent with the results in Fig.~\ref{fig:action_contribution_minimal} and Fig.~\ref{fig:action_contribution_nonminimal}.
We therefore numerically and analytically confirm that the contribution of the potential term is small, less than $1\%$ for reasonable values of the couplings. 
The contribution is much suppressed in the limit of large $\xi$. 
The parameter dependence of large $\xi$ can be well understood since the potential in the Einstein frame depends more strongly on $\xi$ than on $f_a(\phi)$.

On the other hand, it is less trivial that the contribution of the potential term is also suppressed for $\xi\ll 1$.
The origin of the numerical suppression is mainly due to the three-dimensional surface volume of a 3-sphere ($V_{S^3}=2\pi^2$) surrounding the wormhole. It appears in the action Eq.~(\ref{eq:action_with_pot}) as 
\be
\frac{1}{2} \frac{n^2}{V^2_{S^3} f_a^2(\phi) a^6}. 
\ee
The equations of motion give $L_0^2 \propto 1/V_{S^3}$ which is numerically smaller than the expected value from dimensional analysis.

\bibliographystyle{JHEP}
\bibliography{axionbib.bib}
\end{document}